\documentclass{article}
\usepackage{PRIMEarxiv}
\usepackage[draft=true]{minted}
\usepackage{amsmath,amsfonts}
\usepackage{color}
\usepackage{amsmath}
\usepackage{multirow}
\usepackage{graphicx}
\usepackage{textcomp}
\usepackage{ragged2e}
\usepackage{subcaption}
\usepackage{url}
\usepackage[utf8]{inputenc}
\usepackage[T1]{fontenc}
\AtBeginDocument{%
  \providecommand\BibTeX{{%
    \normalfont B\kern-0.5em{\scshape i\kern-0.25em b}\kern-0.8em\TeX}}}

\title{Tensor Ring Optimized Quantum-Enhanced Tensor Neural Networks}

\author{
Debanjan Konar, Dheeraj Peddireddy, Vaneet Aggarwal\\
Purdue Quantum Science and Engineering Institute,\\ 
and School of Industrial Engineering, \\
Purdue University, West Lafayette, USA\\
 \texttt{\{dkonar, dpeddire, vaneet\}@purdue.edu} \\
   \And
 Bijaya K. Panigrahi\\
 Department of Electrical Engineering\\
 Indian Institute of Technology Delhi, New Delhi, India\\
 \texttt{bkpanigrahi@ee.iitd.ac.in} \\
 }
\begin{document}
\maketitle
\begin{abstract}
Quantum machine learning researchers often rely on incorporating Tensor Networks (TN) into Deep Neural Networks (DNN) and variational optimization. However, the standard optimization techniques used for training the contracted trainable weights of each model layer suffer from the correlations and entanglement structure between the model parameters on classical implementations. To address this issue, a multi-layer design of a Tensor Ring optimized variational Quantum learning classifier (Quan-TR) comprising cascading entangling gates replacing the fully connected (dense) layers of a TN is proposed, and it is referred to as Tensor Ring optimized Quantum-enhanced tensor neural Networks (TR-QNet). TR-QNet parameters are optimized through the stochastic gradient descent algorithm on qubit measurements. The proposed TR-QNet is assessed on three distinct datasets, namely Iris, MNIST, and CIFAR-10, to demonstrate the enhanced precision achieved for binary classification. On quantum simulations, the proposed TR-QNet achieves promising accuracy of $94.5\%$, $86.16\%$, and $83.54\%$ on the Iris, MNIST, and CIFAR-10 datasets, respectively. Benchmark studies have been conducted on state-of-the-art quantum and classical implementations of TN models to show the efficacy of the proposed TR-QNet. Moreover, the scalability of TR-QNet highlights its potential for exhibiting in deep learning applications on a large scale. The PyTorch implementation of TR-QNet is available on Github: \url{https://github.com/konar1987/TR-QNet/}.
\end{abstract}


\maketitle

\section{Introduction}
\label{sec:intro}

Deep learning is a very effective and extensively used machine learning method, which has shown great performance in various tasks, including recognition, classification, regression, and clustering~\cite{lathu2020,li2019,he2016,peng2020}. Recent years have witnessed the surge of quantum machine learning~\cite{biamonte2017}, a new computational paradigm that blends quantum computing and machine learning. It employs quantum parallelism and non-classical connections, such as quantum entanglement, to possibly speed up or revolutionize existing classical algorithms~\cite{arute2019}. Importantly, the convergence of these disciplines can result in synergistic improvements and new views on a wide range of difficult challenges~\cite{xiao2022}. Concurrently, combining physics principles and classical machine learning approaches has shown significant promise in tackling quantum computing issues~\cite{qiu2023}. Researchers demonstrated that the trainable weights of neural networks have a strong correlation with many-body wave functions~\cite{gutier2022,pescia2022}. Furthermore, ideas for identifying phase transitions in quantum many-body systems using fully connected artificial neural networks (ANNs) and convolutional neural networks (CNNs) have been examined, with encouraging results~\cite{carras2017,rem2019,zhang2019}.\\
Deep Neural Networks (DNN) have extremely high spatial and temporal complexity levels owing to densely stacked layers containing large-scale matrix multiplications. Hence, DNNs often need several days of training while requiring a considerable amount of memory for inference. Furthermore, substantial weight redundancy in DNNs has been demonstrated~\cite{zeiler2014}, demonstrating the possibility of condensing DNNs while preserving performance. As a result, a variety of compression approaches, including pruning~\cite{molchan2019}, quantization~\cite{yang2019}, and low-rank decomposition~\cite{pan2019}, have been devised. Applying TNs to DNNs to generate TNNs is one of them since TNNs have outstanding potential to approximate the original weights with fewer parameters~\cite{pann2021}, particularly involving the reconstruction of convolutional and fully connected layers using a range of TD formats~\cite{hayashi2019}. However, the scalability of DNN is hindered when a substantial number of neurons are taken into account, thereby restricting the feasible number of layers. This is primarily due to the time-consuming training process and the need for a lot of memory to store the large weight matrices. The accuracy and effectiveness of the DNN model will suffer with an increase in the hidden layers if the parameters for such large-weight matrices are not optimized. Therefore, decreasing the number of model parameters is imperative to maintain accuracy. Nevertheless, the present hardware used to train neural networks significantly restricts their scale and usefulness. These concerns have gained significance due to the imminent approach of physical limitations to impede the progress of performance enhancements in deep classical neural networks. \\
In contemporary times, a correlation has been established between tensor networks (TN) and neural networks, whereby the former serves as an effective ansatz for representing quantum many-body wave functions~\cite{murg2008, orus2014}. As a result, it is possible to substitute tensor networks (TN) for these weights and rely on variational optimization techniques to train them~\cite{stoud2016}. A plethora of TN-based efficient algorithms for classification~\cite{convy2022}, anomaly detection~\cite{fanaee2016,li2011}, segmentation~\cite{konar2023}, and clustering~\cite{stoud2018} have been proposed in recent times. In addition to their capacity for effective expression, TN offers streamlined methodologies for compressing data through tensor factorization techniques~\cite{panaki2021, bahri2019}. For instance, it is possible to significantly reduce the number of parameters in neural network models by retaining only the most significant degrees of freedom and discarding those that exhibit lower correlations. Tensor Neural Networks (TNN)~\cite{konar2023} and Variational Tensor Deep Neural Networks~\cite{hayashi2019,phan2020} are instances of neural networks that rely on tensor network structures to replace the weight tensors of the hidden layers. This is achieved by applying Singular Value Decomposition (SVD) methods. Recent research studies have validated that, despite having a limited parameter space, TNN exhibits superior performance and accuracy compared to conventional ANNs~\cite{panaki2021,novikov2015}. Deep neural network low-rank tensor approximation has been extensively studied in the literature for effective model reduction, low generative error, and high prediction speed~\cite{song2020}.\\
Recently Quantum Neural Networks (QNN) have emerged as a potential contender to circumvent the problems and to facilitate the training of DNN~\cite{konar2016,imri2019,zhao2021, renben2018, konar2020,li2020, konar2022,bausch2020, konar_2022, cerezo2021,abbas2021,beer2020}. Quantum states are mathematical entities of quantum systems compatible with higher-order tensors~\cite{orus2019}. TNNs may thus be utilized as simulators in traditional computers to emulate genuine quantum circuits~\cite{pan2022,huggins2019}. Some particular TNNs can be realized on compact, near-term quantum devices using quantum computing's ultrahigh parallelism~\cite{cohen2016}. Rather than the more broad TN-based quantum circuit modeling paradigm, quantum circuit simulation on TNNs focuses on the functions of TNs as bridges between traditional ANNs and QNNs. 

\subsection{Motivation}
\label{motiv}
Most contemporary TNN advancements involve tensorization solely at the level of hidden layers pertaining to trainable weights~\cite{kossaifi2017,hayashi2019,phan2020, huang2021, jahromi2023}. Training a model typically involves optimizing each layer's contracted trainable weights using established optimization techniques like gradient descent~\cite{konar2023, vaneet2018}. The outcome of this is an adaptable architecture for TNN that can be effectively trained for a substantial quantity of neurons and layers. The variational algorithm employs a method of local gradient-descent, incorporating tensor gradients. This motivates us to propose hybrid TNN models incorporating both tensor and quantum layers. The training algorithm used in our study offers valuable insights into the entanglement structure of trainable weights for fully connected layers of TNN. Nevertheless, it helps to clarify the expressive power of a quantum neural state. 

\subsection{Primary Contributions and Novelty}
\label{contr:novelty}
Considering the neural network's entanglement structure, a novel multi-layer design of a Tensor Ring optimized variational Quantum learning classifier (Quan-TR) with cascading entanglement gates is introduced in the proposed hybrid quantum-enhanced TNN model (TR-QNet).  Furthermore, our TR-QNet model's accuracy and efficiency are evaluated on numerical data and image classification on various datasets. The present study exhibits a tripartite novelty, which can be summarized as follows: 
\begin{enumerate} 
\item Our study presents for the first time a novel quantum-enhanced hybrid tensor neural network (TR-QNet) comprising classical tensor layers followed by quantum layers for data and image classification. The proposed TR-QNet incorporates the novel multi-layer design of Quan-TR, replacing the fully connected softmax layers in TNN, distinguishing it from the state-of-the-art TN models~\cite{hayashi2019,phan2020, huang2021, jahromi2023}. 
\item In addition, the quantum layers (Quan-TR) of the proposed TR-QNet model incorporate a cascading of quantum entangling gates, leading to the elimination of local minima. This is demonstrated by the convergence of the training loss of the proposed TR-QNet model. 
\item Compared with the classical TN model, the binary-class classification accuracy of TR-QNet is improved by $10.53\%$, $7.28\%$ and $12\%$ on the Iris~\cite{iris_2021}, MNIST~\cite{lecun1998} and CIFAR-10~\cite{alexnet2012} datasets, respectively. This approach presents a distinctive and innovative effort towards expediting advancements in resolving computer vision issues through deep quantum learning. 
\end{enumerate}
The subsequent sections of this manuscript are organized in the following manner. Section~\ref{Architecture:TR-QNet} explains the proposed Quantum-Enhanced Tensor Neural Network architecture, which includes an overview of classical Tensor Neural Networks (TNN) and Tensor Ring optimized variational Quantum circuit (Quan-TR). Section~\ref{results:disscuss} contains the datasets, experimental settings, and experimental results. Section~\ref{discuss} elucidates the efficacy of the TR-QNet model and underscores its constraints. Finally, the concluding remarks and future research directions are discussed in Section~\ref{conlud}. 
\emph{Appendix} section provides the convergence of the proposed TR-QNet.
\section{Quantum-Enhanced Tensor Neural Network Architecture}
\label{Architecture:TR-QNet}

\begin{figure*}[!h]
	\centering
	\includegraphics[width=6in]{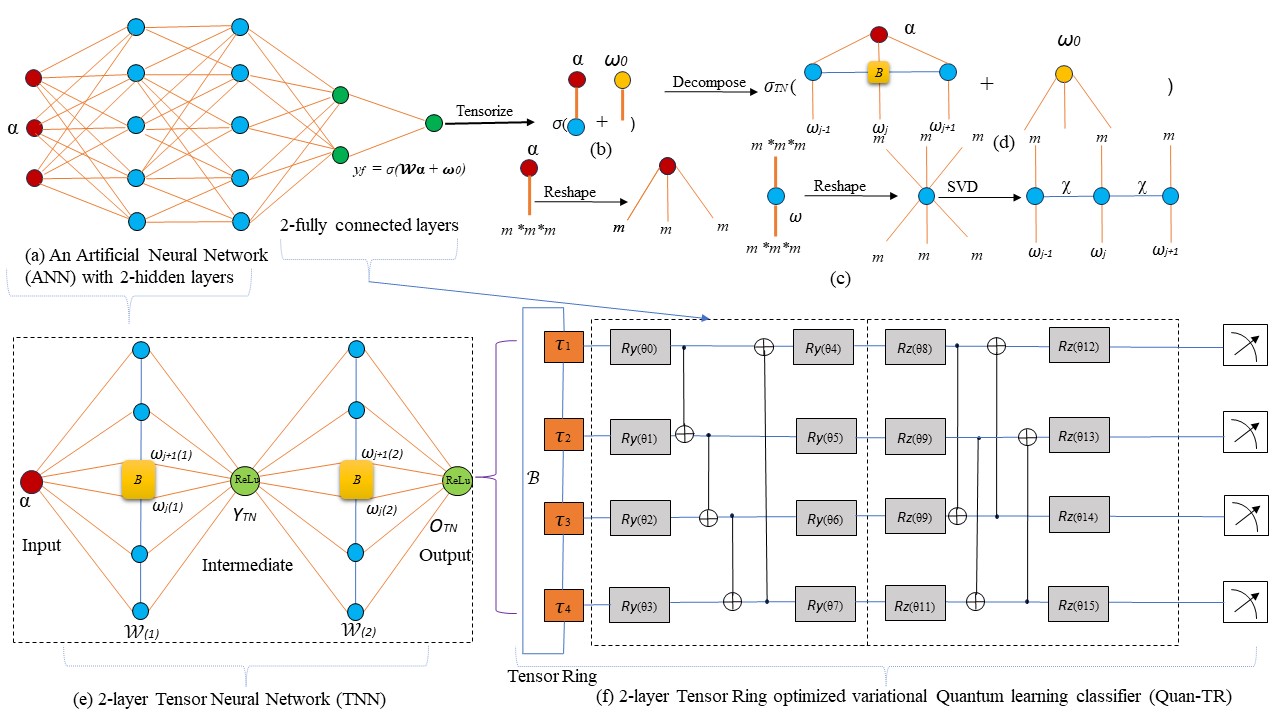} 
	\caption{A Tensor Ring optimized Quantum-enhanced tensor neural Network (TR-QNet) architecture with 4-qubits Quan-TR for tensorizing an (a) ANN with 2 hidden layers and 2 fully connected dense layers. The network prediction, $y_f$, is derived by feeding the model the input feature vector $\alpha$ as $y_f = \sigma(\mathcal{W}\alpha + w_0)$, where $w_0$ is bias vector, and $\sigma$ is the ReLu activation function; (b) the ANN's TN representation using Matrix Product States (MPS) and Matrix Product Operators (MPO); (c) MPO decomposition of the weight matrix $\mathcal{W}$ performing singular value decomposition (SVD) and truncating the inconsequential singular values includes MPO factorization for a matrix $\mathcal{W}_{m^3 \times m^3}$ followed by reshaping $\mathcal{W}$ into a rank-6 tensor and using a suitable SVD, matrix $\mathcal{W}$ may be represented as a 3-site MPO; (d) the resulting TNN with MPO trainable weights; (e) 2-layer Tensor Neural Networks (TNN) tensorizing the ANN using part (b), (c) and (d); and (f) Low-rank Quan-TR component employed in this proposed TR-QNet has three parts: tensor ring encoding ($\tau$), variational learning parameters, and quantum measurement. The cascading CNOT gates are preserved through tensor ring approximation relying on SVD. $\mathcal{R}_y(\theta)$ and $\mathcal{R}_z(\theta)$ are used for data encoding and measurements.}
\label{fig:TR-QNet}
\end{figure*}

The TR-QNet model is a novel proposed framework that combines classical TN and quantum layers (Quan-TR) with tensor ring parameterized inputs and cascading of entangling gates. Recently, the authors \emph{et al.}~\cite{peddi2023} also proposed a similar type of Tensor Ring parameterized Variational Quantum Circuit (TR-VQC). However, the TR-VQC suffers from directly reduced input features due to a few available qubits and the limited entanglement between the parameters, resulting in Barren plateaus~\cite{mcclean2018}. In contrast, our hybrid TR-QNet model exhibits a relationship between tensor neural networks and variational Quantum learning classifiers optimized by a tensor ring structure, as shown in Figure~\ref{fig:TR-QNet}, enabling feeding the full input features through TNN layers with minimal loss of information. The TR-QNet model architecture incorporates a tensor neural network (TNN) with multiple hidden layers. It introduces a multi-layer tensor-ring optimized variational Quantum learning classifier with cascading entangling gates to address quantum entanglement among model parameters efficiently. This approach replaces the conventional soft-max layer typically employed at the end of TN models. A classical pooling layer is incorporated in integrating the TNN model and Quan-TR of the proposed TR-QNet architecture to match the dimension of the input of TNN and the input of Quan-TR. \\
The VQC-based training algorithm resembling DMRG~\cite{white1992} enables a straightforward entanglement of the entanglement spectrum of the Matrix Product Operators' (MPO's)~\cite{panaki2021} trainable weights, thereby facilitating a lucid comprehension of the correlations within the parameters of our TR-QNet model. One can evaluate the MPOs' entanglement structure and capacity as a quantum neural state through standard quantum information measures.

\subsection{Tensor Neural Networks}
\label{tnn}

A Tensor Network Network (TNN) is obtained after the tensorization of an ANN, enabling it to align with the MPO weights' size and dimensions~\cite{murg2008, orus2014}. The hidden layers of the proposed TR-QNet model can reshape into a rank-$d_T$ tensor, possessing a dimension size of $\mathcal{N}_T$, which can subsequently be contracted to form a TN layer. This TN layer comprises six Matrix Product Operators (MPO)~\cite{panaki2021} weights, each having an input size of $m$. Features that cannot be factorized to align with the MPOs in the TN layers are transformed during the preprocessing stage of the training TR-QNet to conform to the input size of the TN layer. A dense trainable layer of size $\mathcal{N}_s \times \mathcal{N}_q$ is added as a connecting layer preceding the Tensor Ring optimized variational Quantum learning classifier (Quan-TR) layer to address the issue of reduction in the size of input data in classical TNN model. The length of the input feature vector of Quan-TR is denoted by $\mathcal{N}_s$, while the output size of the contracted TNN layer is represented by $\mathcal{N}_q$. The contraction of two rank-2 tensors, $\mathcal{S}_{xy}$ and $\mathcal{V}_{yz}$, can be represented diagrammatically by connecting the two tensors along their shared index $y$. Mathematically, the contraction operation is described as follows:
\begin{equation}
\mathcal{T}_{xz} =  Tr(\mathcal{S}_{xy} \mathcal{V}_{yz}) = \sum_y \mathcal{S}_{xy} \mathcal{V}_{yz}
\end{equation}
Here, $Tr$ designates the trace over shared indices $y$.\\
A viable approach to intelligent data compression techniques that rely on TN and MPO decomposition to enhance the representation of weight matrices involves substituting weights with MPOs. The MPO form of the weight matrix of a hidden layer can be derived from the $\mathcal{W}$ matrix by consecutively applying SVD, as demonstrated in Figure~\ref{fig:TR-QNet}. The TN layers comprise a set of trainable weights denoted as $\omega_i$ represented by MPO. A Bond tensor $B_{j,j+1}$ is obtained by contracting a pair of neighboring MPO tensors, $\omega_j$, and $\omega_{j+1}$, along their shared virtual dimension.
Adjusting the input feature vector $\alpha$ to align with the MPO dimensions allows the network's output $\mathcal{O}_{TN}$ to be derived through the contraction of the resulting tensor network. The activation function $\sigma_{TN}$ is applied to the result of a tensor contraction operation~\cite{jahromi2023} as follows.
\begin{equation}
    \mathcal{O}_{TN} = \sigma_{TN}(Tr (\alpha_i,\alpha_j,\alpha_k,\cdots \omega_i,\omega_j,\omega_k, \cdots) + \omega_0)
\end{equation}
Here, the tensor contraction operation between the input tensor $\alpha$ and the weight tensor $\omega$ and $i, j, k,$ etc. represent the tensor indices. It may be noted that the activation function $\sigma_{TN}$ is applied element-wise to the matrix obtained from the tensor contraction, and it cannot be directly applied to individual MPO tensors separately due to the non-linearity introduced by the activation function. In the proposed TR-QNet, one approach involves contracting the features and tensor network layers (MPOs) before applying the activation function and reshaping the resulting tensor to match the inputs of the next layer. This process is repeated until the entire TNN network is contracted. \\

\subsection{Tensor Ring Optimized Variational Quantum Learning Classifier} 
\label{tr:vqc}

The proposed Tensor Ring optimized variational Quantum learning classifier (Quan-TR) with Tensor Neural Networks is a hybrid classical-quantum algorithm combining tensor network elements and variational quantum circuits for data and image classification. The proposed Quan-TR introduces a multi-layer Tensor Ring optimized variational Quantum learning classifier with cascading entangling gates to address quantum entanglement among model parameters efficiently, which is the major distinction with our Tensor Ring Parameterized Variational Quantum Circuit (TR-VQC)~\cite{peddi2023}. The proposed Quan-TR framework consists of three main components: tensor ring encoding, variational learning parameters, and measurement. tensor ring encoding represents the quantum states in a compressed format. It leverages the tensor ring structure, a tensor network with a specific hierarchical ring-like connectivity pattern. The tensor ring approximation uses SVD to compress the quantum states while preserving important features. This approximation allows for efficient representation and manipulation of quantum states within Quan-TR. In the proposed Quan-TR, single-qubit rotation gates $\mathcal{R}_y(\theta)$, and $\mathcal{R}_z(\theta)$ are used to represent rotations along the Y and Z axes, respectively. These rotation angles ($\theta$) are learned during training to find the optimal values that minimize the objective function. By combining tensor ring encoding with variational learning parameters and measurement, the proposed TR-QNet architecture enables the training of quantum circuits for data and image classification. In our Quan-TR framework, the tensor ring parametrization represents a quantum state $|\psi\rangle$ using $V$ tensors, each with bond dimension $\mathcal{B}$, denoted by $\tau (\upsilon)$ as follows~\cite{vaneet2018}: 
\begin{equation}
\begin{split}
|\psi\rangle = \sum_{x_i}^{x_V} \sum_{y_i}^{y_V} \tau(1)^{y_1}_{x_V x_1}\tau(2)^{y_2}_{x_1 x_2}\cdots \tau(V)^{y_V}_{x_{V-1} x_V}|x_i, x_2, \cdots x_V\rangle
\end{split}
\end{equation}
The physical indices $x_\upsilon \in \{0, 1\}$ span the $2^V$-dimensional Hilbert space, while the bond indices $y_\upsilon \in \{1, \cdots N_\upsilon \}$, control the maximum amount of entanglement captured by the tensor ring, also known as tensor rank. A tensor ring parametrization of a 4-qubit state is illustrated in Figure~\ref{fig:TR-QNet}. In Quan-TR, each $\tau (\upsilon)$ in the ring represents a tensor of dimension $\mathcal{B} \times \mathcal{B} \times \mathcal{X}$, signifying the connections between the tensors in the tensor ring. The tensor, $\tau (\upsilon)$, has three indices, two of which have a bond dimension $\mathcal{B}$, and the third index has a dimension $\mathcal{X}$. Subsequently, the input characteristics are encoded through the utilization of single qubit rotation gates ($\mathcal{R}_y (\theta)$), which preserves the tensor ring configuration. The fundamental element of the parametrized circuit in every layer of the proposed Quan-TR model is the cascading entanglement of qubits, which is subsequently followed by parametrized single qubit rotations. The two-qubit gates, such as the CNOT gate, do not preserve the tensor ring representation. An approximation technique based on singular value thresholding is employed for this gate to address this issue. The tensor ring structure facilitates the computation of 2-qubit gates for adjacent qubits. By employing a cascading configuration of the tensor ring, executing a CNOT operation from the ultimate qubit to the initial qubit becomes feasible. It is worth noting that using the tensor ring format allows for the utilization of the same rank $d_q$ in each decomposition, which may not be feasible with the conventional Matrix Product State (MPS) format. By employing this approximation, all calculations for the forward pass exhibit linearity concerning the number of gates.\\
We develop a universal TR-QNet model that uses the intrinsic probabilistic behavior of qubit measurements to classify images using a hybrid classical-quantum framework. The aspects of variational quantum learning classifier concerning encoding, variational, and measurement are all accomplished within the implementation of Quan-TR. Single-qubit rotation gate, $\mathcal{R}_y(\theta)$, is employed to encode rotations along the $Y$-axes in the encoding section. 
Quantum bits (qubits) represent the input state of VQC in the proposed TR-QNet as
\begin{equation}
    |\psi(\theta)\rangle = (\cos \theta |0\rangle + \sin \theta |1\rangle)|\mathcal{O}_{TN}\rangle\ .
\end{equation}
In the VQC of the TR-QNet model, the quantum states $|\psi(\theta)\rangle$ correspond to the quantum encoding of the classical inputs $\mathcal{O}_{TN}$ from the classical layer of TNN. 
The Tensor Ring parameterized quantum circuit (Quan-TR) is dense and constitutes parametrized single qubit gates with CNOT gates to entangle quantum states from each qubit. To encode phase information, the dressed quantum layer of TR-QNet uses the rotation gates $\mathcal{R}_y$ and $\mathcal{R}_z$. Complementary quantum states are created with the help of the CNOT gate. In the Bloch sphere projection, the $\mathcal{R}_y(\theta)$ and $\mathcal{R}_z(\theta)$ gates represent the following single-qubit rotations about the $Y$ and $Z$-axes, respectively, as follows:
\begin{equation}
    \mathcal{R}_y(\theta)= \exp{(-jY\theta/2)}=\left[ {{\begin{array}{*{20}c}
		{\cos\theta/2 } \hfill & {-\sin\theta/2 } \hfill \\
		{\sin\theta/2 } \hfill & {\cos\theta/2 } \hfill \\
		\end{array} }} \right]
\end{equation}
and 
\begin{equation}
    \mathcal{R}_z(\theta)= \exp{(-jZ\theta/2)}=\left[ {{\begin{array}{*{20}c}
		{\exp{(-j\theta/2)}} \hfill & ~~0 \hfill \\
		~~0 \hfill & 	{\exp{(j\theta/2)}} \hfill \\
		\end{array} }} \right]\ .
\end{equation}
To perform the one-qubit rotation, we contract the $2 \times 2$ unitary rotation matrix $\mathcal{R}$ with the original tensor $\tau (\upsilon)$, and the resulting tensor $\tau' (\upsilon)$ represents the rotated state of the $\upsilon^{th}$ qubit as follows:
\begin{equation}
\begin{split}
   \tau'(\upsilon)^{y'_\upsilon}_{x_{\upsilon-1}x_\upsilon} = \sum_{y_\upsilon} \mathcal{R}_{y'_\upsilon y_\upsilon}\tau(\upsilon)^{y_\upsilon}_{x_{\upsilon-1}x_\upsilon}
\end{split}
\end{equation}
To perform a two-qubit gate transformation on qubits $\upsilon$ and ($\upsilon +1$) in the proposed Quan-TR, the tensor network needs to be transformed into an orthogonal form centred around the qubits of interest $\upsilon$ and ($\upsilon +1$) before applying the gate operation. The shared bond index is contracted between the tensors $\tau (\upsilon)$ and $\tau (\upsilon+1)$ to create a new tensor as follows:
\begin{equation}
    \mathcal{M}^{y_\upsilon y_{\upsilon+1}}_{x_{\upsilon-1} x_{\upsilon+1}}= \sum_{x_\upsilon} \tau(\upsilon)^{y_\upsilon}_{x_{\upsilon-1}x_\upsilon} \tau(\upsilon +1)^{y_{\upsilon+1}}_{x_{\upsilon}x_{\upsilon +1}}
    \label{eq:tensor_contr}
\end{equation}
To apply the two-qubit gate $\mathcal{U}$ on the two-qubit tensor computed from Equation~\ref{eq:tensor_contr}, we reshape the gate $\mathcal{U}$ into an operator acting on the joint state of qubits $\upsilon$ and ($\upsilon +1)$. 
\begin{equation}
(\mathcal{M'})^{y_\upsilon y_{\upsilon+1}}_{x_{\upsilon-1} x_{\upsilon+1}} = \sum_{y_{\upsilon} y_{\upsilon+1}} \mathcal{U}_{y'_{\upsilon} y'_{\upsilon+1}y_{\upsilon} y_{\upsilon+1}}\mathcal{M}^{y_\upsilon y_{\upsilon+1}}_{x_{\upsilon-1} x_{\upsilon+1}}
\end{equation}
We perform SVD on the resultant tensor $\mathcal{M'}$ on reshaping it as $(y'_{\upsilon}+ x_{\upsilon-1}) \times (y'_{\upsilon +1}+ x_{\upsilon+1})$ as follows:
\begin{equation}
(\mathcal{M'})^{y_\upsilon y_{\upsilon+1}}_{x_{\upsilon-1} x_{\upsilon+1}} = \sum_{x_{\upsilon}}\mathcal{P}^{y'_\upsilon}_{x_{\upsilon-1}x_\upsilon} \mathcal{S}_{x_\upsilon}\mathcal{Q}^{y'_{\upsilon+1}}_{x_{\upsilon}x_{\upsilon +1}}
\end{equation}
Here, $\mathcal{P}$ and $\mathcal{Q}$ comprise orthogonal vectors, $\mathcal{S}_{x_\upsilon}$ is composed of singular values of matrix $\mathcal{M'}$. The matrix has $2N$ singular values irrespective of the two-qubit gate structure, where $N$ denotes the bond dimension of the tensor ring. We then truncate the $\mathcal{S}_{x_\upsilon}$ matrix to keep only the $N$ largest singular values, and the resulting matrix is denoted by $\mathcal{S'}_{x_\upsilon}$. $\mathcal{P}$ and $\mathcal{Q}$ are truncated only to keep the orthogonal vectors corresponding to the $N$ largest singular values. 
\begin{equation}
\tau'(\upsilon)^{y_\upsilon}_{x_{\upsilon-1}x_\upsilon}=\mathcal{P}^{y_\upsilon}_{x_{\upsilon-1}x_\upsilon}\mathcal{S'}_{x_\upsilon}
\end{equation}
and 
\begin{equation}
\tau'(\upsilon+1)^{y_{\upsilon+1}}_{x_{\upsilon+1}x_\upsilon}=\mathcal{Q}^{y_{\upsilon+1}}_{x_{\upsilon}x_{\upsilon+1}}
\end{equation}
The preprocessed data from the TNN layer, denoted as $\mathcal{O}_{TN}^i$, is transformed into a quantum state represented by $|\psi(\mathcal{O}_{TN}^i)\rangle$. Subsequently, the quantum state undergoes processing through Quan-TR with parameters $\mathcal{U}(\theta_1, \theta_2,\cdots, \theta_n)$. Finally, by performing measurements on particular qubits through the use of the Pauli-Z basis, we obtain a collection of outputs denoted as ${\lambda_j}$ along with their corresponding probabilities as follows:
\begin{equation}
\mathcal{Z}_{ij} = \langle \psi (\mathcal{O}_{TN}^i)|\mathcal{U}^\dagger (\theta)|\lambda_j\rangle \langle \lambda_j|\mathcal{U} (\theta)|\psi (\mathcal{O}_{TN}^i)\rangle
\end{equation}
where, complete operation is $\mathcal{U} (\theta)$ is defined as 
\begin{equation}
\mathcal{U} (\theta) = \mathcal{U}_n (\theta_n)\mathcal{U}_{n-1} (\theta_{n-1}) \cdots \mathcal{U}_1 (\theta_1).
\end{equation}
The loss function, $\mathcal{L}(\theta)$, can be defined as follows considering the input quantum state as $|0\rangle^\mathcal{N}_q$.
\begin{equation}
\begin{split}
   \mathcal{L}(\theta) = f(y_j (\theta), t_j)=\mathcal{Z}(y_j(\theta) \neq t_j) =\\\sum_{j}^{\mathcal{N}_q} f((\langle 0|\psi^\dagger (\mathcal{O}_{TN}^j)\mathcal{U}^\dagger (\theta) y_j \mathcal{U} (\theta)\psi (\mathcal{O}_{TN}^j)|0\rangle), t_j)
   \end{split}
\end{equation}
where, $y_j(\theta) \in \{\overline{\lambda_j}\}$ and $t_j$ corresponds to a target output. In order to train the proposed Quan-TR model, the gradient of the loss function is evaluated as follows:
\begin{equation}
\begin{split}
\frac{\delta \mathcal{L}(\theta)}{\delta \theta_j} = \langle 0|\psi^\dagger (\mathcal{O}_{TN}^j)\frac{\delta\mathcal{U}^\dagger (\theta)}{\delta \theta_j} y_j \mathcal{U} (\theta)\psi (\mathcal{O}_{TN}^j)|0\rangle +  \langle 0|\psi^\dagger (\mathcal{O}_{TN}^j)\mathcal{U}^\dagger (\theta) y_j \frac{\delta\mathcal{U} (\theta)}{\delta \theta_j}\psi (\mathcal{O}_{TN}^j)|0\rangle = \\
\langle 0|\psi^\dagger (\mathcal{O}_{TN}^j)\mathcal{U}_1^\dagger (\theta_1)\cdot\frac{\delta\mathcal{U}_j^\dagger (\theta_j)}{\delta \theta_j}\cdot \mathcal{U^\dagger}_n (\theta_n)y_j \mathcal{U} (\theta)\psi (\mathcal{O}_{TN}^j)|0\rangle + \\ \langle 0|\psi^\dagger (\mathcal{O}_{TN}^j)\mathcal{U}^\dagger (\theta) y_j \mathcal{U}_n (\theta_n)\cdot\frac{\delta\mathcal{U}_j (\theta_j)}{\delta \theta_j}\cdot\mathcal{U}_1 (\theta_1)\psi (\mathcal{O}_{TN}^j)|0\rangle\\
= \langle 0|\psi^\dagger (\mathcal{O}_{TN}^j)\mathcal{U}_{-}^\dagger[i\psi_j] \mathcal{U}_{+}^\dagger y_j \mathcal{U} (\theta)\psi (\mathcal{O}_{TN}^j)|0\rangle + \langle 0|\psi^\dagger (\mathcal{O}_{TN}^j)\mathcal{U}^\dagger (\theta) y_j \mathcal{U}_{+}[-i\psi_j]\mathcal{U}_{-}\psi (\mathcal{O}_{TN}^j)|0\rangle
\end{split}
\end{equation}
where, $\mathcal{U}_j (\theta_j) = e^{-i\theta_j\psi (\theta^j)}$.\\
However, due to NISQ's limitations, classical simulators are now being utilized to optimize and update parameters and feed them back to TNN and Quan-TR separately until convergence conditions are reached. Hence, we have used cross-entropy loss to update the parameters.
The loss function ($\mathcal{L}_{\theta}$) is derived with the hyper-parameters $\theta$ of the proposed TR-QNet model as
\begin{equation}
\begin{split}
   \operatorname*{argmin}_\theta \mathcal{L}_{(\theta)} = \sum_{j}^{\mathcal{N}_q} [t_j \log \overline{f}(\mathcal{O}_{TN}^j) + (1-t_j) \log \{1-\overline{f}(\mathcal{O}_{TN}^j)\}]\ .
   \end{split}
   \label{eq:loss_func}
\end{equation}
where, $\overline{f}(\mathcal{O}_{TN}^j, \theta)$ can defined for binary optimization problem as follows.
\begin{equation}
\overline{f}(\mathcal{O}_{TN}^j, \theta) = 
\begin{cases}
1, & \text{if } f(\mathcal{O}_{TN}^j, \theta) >0\\
-1, & \text{otherwise}
\end{cases}
\end{equation}
The DMRG-like sweeping technique~\cite{white1992} for training the TNN uses a stochastic gradient-based optimization strategy in which a gradient descent step with a learning rate updates the local bond tensors $Bj,j+1$ towards a global minimum of the loss function. 
In order to update the weights in TNN, a gradient of the bond tensors with respect to the loss, $Bj,j+1$, is obtained by defining $\overline{f}(\mathcal{O}_{TN}^j)=TB$, where $T$ represents the contraction of every tensor in the TNN other than the bond tensor $B$.

\section{Results}
\label{results:disscuss}

\subsection{Data Sets}
\label{results:dataset}
The Iris dataset~\cite{iris_2021} is often used as a benchmark to evaluate the performance of different machine learning algorithms. The Iris dataset contains a total of $150$ samples, and each sample has four features: sepal length, sepal width, petal length, and petal width. We extracted three distinct binary data sets from the original Iris dataset. We added $80\%$ samples to each training subset and the remaining $20\%$ samples for each class as a test data set. \\
Researchers studying computer vision often rely on the MNIST dataset~\cite{lecun1998} as a benchmark for Artificial Neural Networks. The MNIST dataset has $70,000$ $28 \times 28$ grayscale images ($60,000$ for training and $10,000$ for testing), divided into $10$ classes and each containing $7,000$ images.\\
The CIFAR-10~\cite{alexnet2012} dataset comprises a total of $60,000$ images in $10$ categories ($6000$ for each class), with $32 \times 32$ colour images including $50,000$ training images and $10000$ test images. The tests, however, resize and transform CIFAR-10 images into $28 \times 28$ times their original size grey-scale images. 
\\However, owing to the limited qubit available at the NISQ processor, we perform binary classification jobs using this batch of images with values $0$ or $9$, $1$ or $8$, $2$ or $7$, $3$ or $6$, and $4$ or $5$  and multi-class classification with values $0$, $1$ or $9$, $2$, $4$ or $5$ and $3$, $6$ or $7$. We had to restrict our datasets to two randomly selected classes in our investigations since the Qiskit Quantum Simulator only has access to a few qubits.

\subsection{Experimental Settings}
\label{experiment:result}

We compute the original input data sets' mean and variance. The data sets are then normalized using the zero-mean normalization procedure to have a zero mean and unit variance before feeding into the TNN. The proposed TR-QNet comprises TN layers with several trainable MPO tensors, and a stochastic gradient-based algorithm has been employed to train the MPOs~\cite{panaki2021}, relying on a DMRG-like technique. The prior tensor gradient technique is appropriate for TNN models fully composed of TN layers. However, it is neither effective nor adaptable in models with hybrid architectures that combine TN layers and VQC. \\
We employed automated differentiation techniques~\cite{baydin2018} and a classical back-propagation algorithm to determine the gradient of the TNN trainable weights as our TR-QNet is a feed-forward hybrid neural network combining TN and quantum layers. We have used \emph{TensorLy-Torch} library to compute the automatic differentiation of TN layers in PyTorch settings. However, being a hybrid classical-quantum framework, the classical TNN model is simulated on classical hardware and Quan-TR on the Qiskit simulator. The weights of the TN layers are updated using a layer-by-layer approach. The intermediate dense layer has merely been included to make up for the size mismatch between the features in the last TN layer and Quan-TR, and it is not trainable. Each $6$ MPO trainable weight on the TN layers has virtual dimension $V$ and a \emph{ReLu} activation function. The last layer of Quan-TR in the output chain is a dense layer with softmax activation, which outputs vectors that are one-hot encoded (OHE) and contain the predicted probabilities for the desired number of labels. We set up the initial $V$ qubit state as $|00\cdots 0\rangle$, which is afterward transformed into a Tensor Ring (TR) representation since $\tau (\epsilon)$ is a $\mathcal{B} \times \mathcal{B} \times 2$ tensor with only ($0, 0, 0$)$^{th}$ element as $1$ and rest as $0's$. Our Quan-TR is repeated $r$ times to illustrate the depth of Quan-TR. The tensor ring rank in Quan-TR is set to $d_q=4$ for all tests. \\
Experiments have been carried out using the varying numbers of qubits ($4, 6, 8, 10, 12$) and number of TN and Quan-TR layers on an Nvidia Tesla $V100-SXM2$ GPU Cluster with $32$ GB of memory and $640$ Tensor cores with $8$ cores of Intel(R) Xeon(R) CPU E5-2683 v4@2.1GHz. In the case of image classification, $784$ input features ($28 \times 28$) from the input images are received at the input layer of the proposed TR-QNet. With a maximum of $25$ epochs, Quan-TR layers of the proposed TR-QNet model are rigorously trained using the Adam optimizer with an initial learning rate of $0.01$ and weight decay ($\delta$) of $0$. Figure~\ref{fig:Train_loss} shows the convergence of loss during training of the proposed TR-QNet model varying number of qubits and TN and Quan-TR layers with $5$-fold cross-validation. In the Iris data classification experiments, the proposed Quan-TR is provided with the four feature vectors ($\mathcal{N}_q =4)$ for training from the previous TNN layer through the dense layer with batch size $4$. We chose three measures at random to represent the three classes of the dataset out of the $2^4$ available measurements acquired from Quan-TR. To further transform selected measurements into class probabilities, we employ the sigmoid activation function (Softmax) and the cross-entropy loss function as given in Equation~\ref{eq:loss_func}. However, in the case of MNIST and CIFAR-10 datasets, for binary classification, we choose the final measurements $|00 \cdots 0\rangle$ and $|11\cdots1\rangle$ as the output values and batch size $32$, where multiple readouts need to feed the results to TR-QNet. 

\subsection{Experimental Results}
\label{result:exp}

Extensive experiments have been conducted using large sets of Iris~\cite{iris_2021}, MNIST~\cite{lecun1998}, and CIFAR-10~\cite{alexnet2012} datasets with varying numbers of qubit count $4$, $6$, $8$, $10$, and $12$ and tensor ring ranks ($d_q$) of $2$, $3$, and $4$ as provided in Table~\ref{tab1}. However, it has been found from the experimental data for the Iris dataset that the optimal result is found for $4$ qubits Quan-TR model with tensor ring rank of $4$ as reported in Table~\ref{tab1}. It is noteworthy that the proposed TR-QNet, and its quantum counterparts, namely, Variational Quantum Tensor Networks classifier (VQTN)~\cite{huang2021}, Quantum Convolutional Neural Networks (QCNN)~\cite{cong2019}, Tensor Ring parametrized Variational Quantum Circuit (TR-VQC)~\cite{peddi2023}, and fully classically simulated Variational Tensor Neural Network (VTNN)~\cite{jahromi2023} are trained on the binary and ternary pair of classes from the datasets.\\
In order to illustrate the resilience of the proposed model over the quantum counterpart and classical tensor neural network-based models, unforeseen test images on the Iris~\cite{iris_2021}, MNIST~\cite{lecun1998}, and CIFAR-10~\cite{alexnet2012} datasets are used for evaluation. The training loss curves for the proposed TR-QNet model are demonstrated on the Iris~\cite{iris_2021} in Figure~\ref{fig:Train_loss}. The convergence analysis of the proposed TR-QNet is also provided in the \emph{Appendix}.  
\begin{figure}[htbp]
 \centering
 \subcaptionbox{}{\includegraphics[width=2.5in]{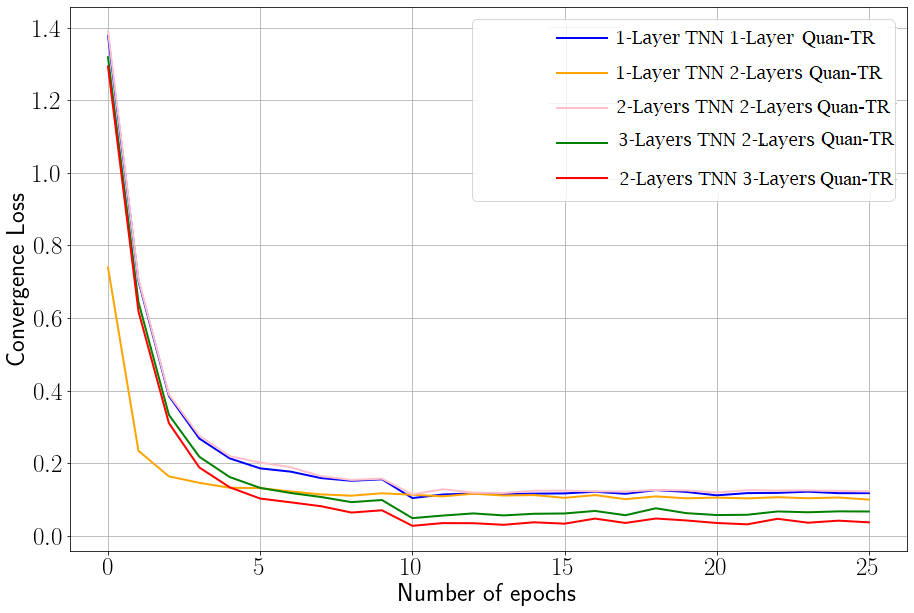}}
 \subcaptionbox{}{\includegraphics[width=2.5in]{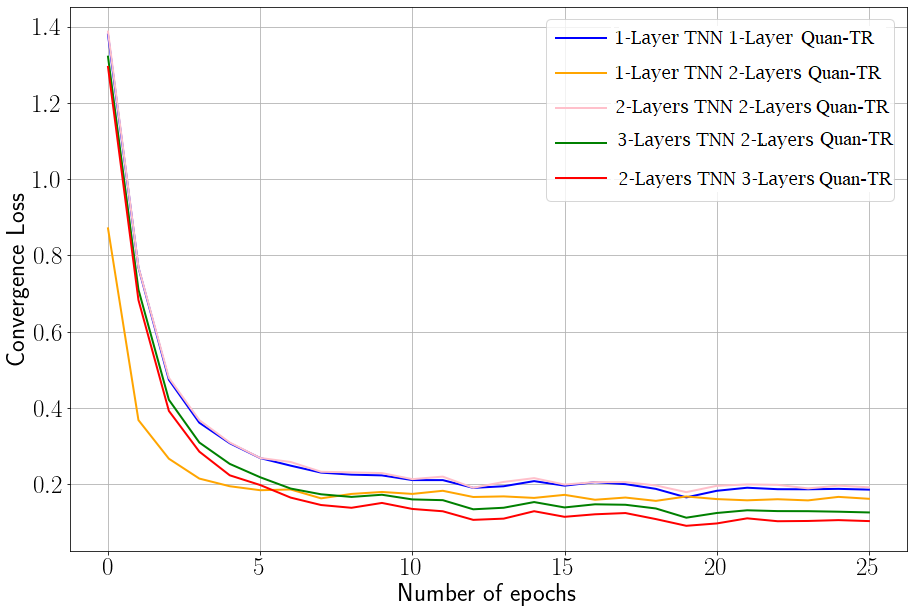}}
 \subcaptionbox{}{\includegraphics[width=2.5in]{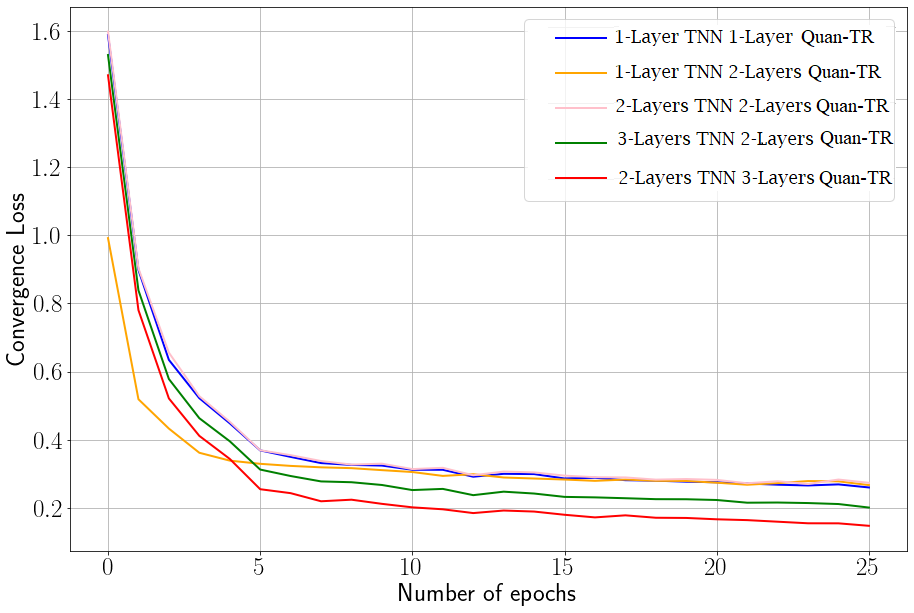}}
 \subcaptionbox{}{\includegraphics[width=2.5in]{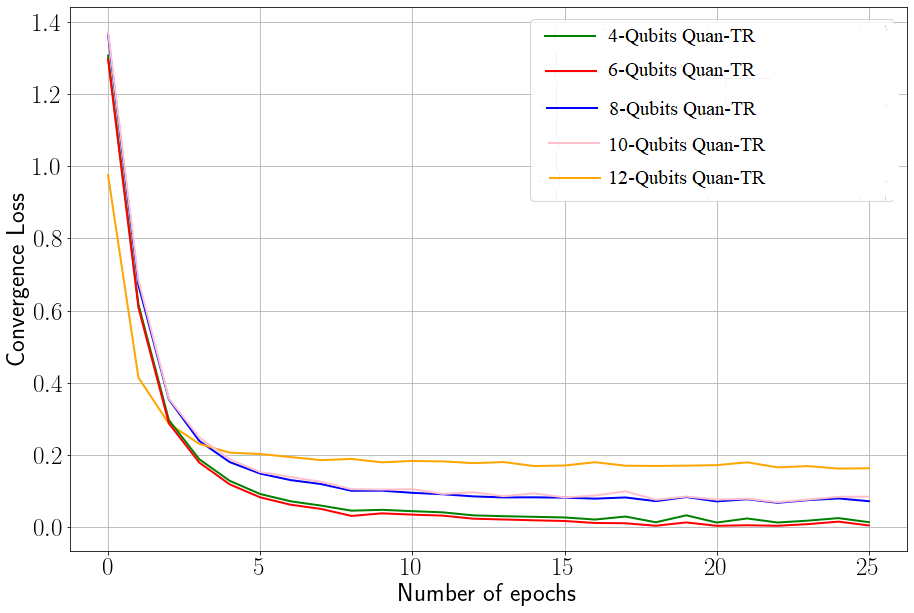}}
 \caption{TR-QNet training loss is reported on a $4$ qubits systems with varying layers of TNN and Quan-TR layers for randomly selected binary classes (a) 1 or 2, (b) 2 or 3 and (c) 1 or 3, (d) varying with qubits ($4, 6, 8, 10$, and $12$ on Iris dataset~\cite{iris_2021}.}
 \label{fig:Train_loss}
 \end{figure}
Table~\ref{tab2} summarizes the numerical results obtained using our TR-QNet using $4$, $6$ and $8$ number of qubits, VQTN~\cite{huang2021}, and QCNN~\cite{cong2019} and TR-VQC~\cite{peddi2023} using $4$ (Iris dataset) and $8$ qubits (MNSIT and CIFAR-10 datasets) and fully classically simulated VTNN~\cite{jahromi2023} on Iris, MNSIT and CIFAR-10 datasets. It has been observed from the experimental results reported in Table~\ref{tab2} that optimal accuracy has been achieved for class $2$ or $3$ in most cases of the Iris dataset. On the contrary, in the case of MNIST and CIFAR-10 datasets, class $3$ or $6$ reports optimal accuracy for most of the models discussed in the manuscript. Our TR-QNet achieves promising accuracy of $94.5\%$, $86.16\%$, and $83.54\%$ with $4$ qubits on the Iris and with $6$ qubits on the MNIST and CIFAR-10 datasets, respectively. However, in the case of multi-class classification, despite TR-QNet's low accuracy, as provided in Table~\ref{tab3}, it outperforms VQTN and VTNN. It may be noted that TR-VQC and QCNN are not feasible for multi-class classification owing to the limitations of their framework. In addition, we use a $\gamma = 0.05$ significant threshold for a two-sided paired Wilcoxon signed-rank test~\cite{conover1999} to demonstrate the effectiveness of the proposed TR-QNet model over other methods. It is evident from the two-sided paired Wilcoxon signed-rank test that the proposed TR-QNet model yields statistically significant results using $4$ and $6$ qubits Quan-TR for Iris data and image (MNIST and CIFAR-10) classification, respectively. This is primarily owing to the limited Iris data feature demanding fewer qubits, whereas the larger image size requires more qubits. Further increasing to $8, 10$, and $12$ qubits resulted in a substantial decrease in accuracy for the proposed TR-QNet and the other methods, probably as a result of over-parametrization~\cite{larocca2023} and barren plateaus~\cite{mcclean2018}. 
\begin{table}[!h]
\footnotesize
	\begin{center}
		\caption{Comparative analysis of the proposed 2-2 layers TR-QNet with varying number of qubits and tensor ranks ($d_q$) on Iris dataset}
		\begin{tabular}{p{15pt}p{0pt}p{20pt}p{20pt}p{20pt}p{0pt}p{20pt}p{20pt}p{20pt}p{0pt}p{20pt}p{20pt}p{20pt}}
			\hline
			\multirow{2}{*}{\centering{\textbf{Qubit}}} & &
			\multicolumn{3}{p{60pt}}{\centering{\textbf{$d_q$=2}}} & &
            \multicolumn{3}{p{60pt}}{\centering{\textbf{$d_q$=4}}} & &
			\multicolumn{3}{p{60pt}}{\centering{\textbf{$d_q$=6}}} \\
			\cline{3-5}
			\cline{7-9}
			\cline{11-13}
			  & & $\textbf{1 or 2}$ & $\textbf{2 or 3}$ & $\textbf{1 or 3}$ & & $\textbf{1 or 2}$ & $\textbf{2 or 3}$ & $\textbf{1 or 3}$ & & $\textbf{1 or 2}$ & $\textbf{2 or 3}$ & $\textbf{1 or 3}$\\
            \cline{1-1}
			\cline{3-5}
			\cline{7-9}
			\cline{11-13}
			4 & & $0.919$ & $0.875$ & $0.891$ & & $0.941$ & $0.939$ & $0.955$ & & $0.939$ & $0.911$ & $0.924$\\	
                6 & & $0.801$ & $0.794$ & $0.789$ & & $0.882$ & $0.865$ & $0.880$ & & $0.829$ & $0.817$ & $0.829$\\
			8 & & $0.782$ & $0.765$ & $0.770$ & & $0.787$ & $0.780$ & $0.788$ & & $0.782$ & $0.757$ & $0.772$\\
            10 & & $0.728$ & $0.765$ & $0.760$ & & $0.773$ & $0.743$ & $0.760$ & & $0.769$ & $0.750$ & $0.763$\\
            12 &  & $0.628$ & $0.605$ & $0.616$ & &  $0.673$ & $0.643$ & $0.690$ & & $0.629$ & $0.617$ & $0.609$\\
          \hline
		    \label{tab1}
		\end{tabular}
	\end{center}
\end{table}
\begin{table*}[htbp]
\footnotesize
	\begin{center}
		\caption{Mean accuracy of the proposed TR-QNet with VQTN~\cite{huang2021}, TR-VQC~\cite{peddi2023}, QCNN~\cite{cong2019}, and fully classically simulated VTNN~\cite{jahromi2023} on the test Iris~\cite{iris_2021}, MNIST~\cite{lecun1998} and CIFAR-10~\cite{alexnet2012} datasets [The bold values sheds light to the two-sided paired Wilcoxon signed-rank test data~\cite{conover1999}]}
		\begin{tabular}{p{25pt}p{25pt}p{1pt}p{15pt}p{15pt}p{15pt}p{1pt}p{15pt}p{15pt}p{15pt}p{15pt}p{15pt}p{1pt}p{15pt}p{15pt}p{15pt}p{15pt}p{15pt}}
			\hline
			\multirow{2}{*}{\centering{\textbf{Model}}} &
			\multirow{2}{*}{\centering{\textbf{Qubits}}} & &
			\multicolumn{3}{p{45pt}}{\centering{\textbf{Iris}}} & &
			\multicolumn{5}{p{60pt}}{\centering{\textbf{MNIST}}} & &
			\multicolumn{5}{p{60pt}}{\centering{\textbf{CIFAR-10}}}\\
			  \cline{4-6}
			\cline{8-12}
			\cline{14-18}
			 & & & $\textbf{1 or 2}$ & $\textbf{2 or 3}$ & $\textbf{1 or 3}$ & & $\textbf{0 or 9}$ & $\textbf{1 or 8}$ & $\textbf{2 or 7}$ & $\textbf{3 or 6}$ & $\textbf{4 or 5}$ & & $\textbf{0 or 9}$ & $\textbf{1 or 8}$ & $\textbf{2 or 7}$ & $\textbf{3 or 6}$ & $\textbf{4 or 5}$ \\
			 \cline{1-2}
			 \cline{4-6}
		    \cline{8-12}
			\cline{14-18}
			\multirow{3}{*}{\textbf{TR-QNet}} & 4 & & $\textbf{0.941}$ & $\textbf{0.939}$ & $\textbf{0.955}$ & & $0.817$ & $\textbf{0.828}$ & $0.869$ & $0.850$ & $0.809$ & & $0.758$ & $0.803$ & $0.747$ & $0.798$ & $0.771$ \\
   &  6 & & $0.882$ & $0.865$ & $0.880$ & & $\textbf{0.828}$ & $\textbf{0.836}$ & $\textbf{0.891}$ & $\textbf{0.863}$ & $\textbf{0.870}$ & & $\textbf{0.819}$ & $\textbf{0.849}$ & $\textbf{0.833}$ & $\textbf{0.857}$ & $\textbf{0.809}$ \\
			&  8 & & $0.787$ & $0.780$ & $0.788$ & & $0.667$ & $0.684$ & $0.669$ & $0.650$ & $0.671$ & & $0.658$ & $0.603$ & $0.647$ & $0.618$ & $0.624$ \\
    \hline
    	\multirow{1}{*}{\textbf{VQTN}} &  4/6 & & $\textbf{0.924}$ & $0.905$ & $0.911$ & & $0.813$ & $0.806$ & $0.829$ & $0.811$ & $0.823$ & & $0.788$ & $0.794$ & $0.776$ & $0.745$ & $0.763$ \\
     \hline
    	\multirow{1}{*}{\textbf{QCNN}} &  4/6 & & $0.871$ & $0.852$ & $0.861$ & & $0.772$ & $0.736$ & $0.740$ & $0.742$ & $0.755$ & & $0.721$ & $0.714$ & $0.717$ & $0.732$ & $0.746$ \\
   \hline
   	\multirow{1}{*}{\textbf{TR-VQC}} &  4/6 & & $0.853$ & $0.849$ & $0.829$ & & $0.803$ & $0.799$ & $0.802$ & $0.789$ & $0.790$ & & $0.767$ & $0.759$ & $0.761$ & $0.753$ & $0.747$ \\
    \hline
   	\multirow{1}{*}{\textbf{VTNN}} & N/A & & $0.838$ & $0.839$ & $0.842$ & & $0.797$ & $0.788$ & $0.798$ & $0.778$ & $0.780$ & & $0.701$ & $0.698$ & $0.734$ & $0.727$ & $0.715$ \\
   \hline
\label{tab2}
\end{tabular}
\end{center}
\end{table*}
\begin{table}[!h]
\footnotesize
	\begin{center}
		\caption{Mean accuracy of the proposed TR-QNet with tensor ranks ($d_q=4$), VQTN~\cite{huang2021} and VTNN~\cite{jahromi2023} for multi-class (3-class) classification [The bold numbers provide information about the two-sided paired Wilcoxon signed-rank test data]}
		\begin{tabular}{p{30pt}p{15pt}p{0pt}p{20pt}p{0pt}p{30pt}p{30pt}p{30pt}p{0pt}p{30pt}p{30pt}p{30pt}}
			\hline
		 & & &
			\multicolumn{1}{p{20pt}}{\centering{\textbf{Iris}}} & &
            \multicolumn{3}{p{90pt}}{\centering{\textbf{MNIST}}} & &
			\multicolumn{3}{p{90pt}}{\centering{\textbf{CIFAR-10}}} \\
			\cline{3-4}
			\cline{6-8}
			\cline{10-12}
			  \multirow{1}{*}{\centering{\textbf{Model}}} &
			\multirow{1}{*}{\centering{\textbf{Qubit}}} & & $\textbf{1, 2 or 3}$ & & $\textbf{0, 1 or 9}$ & $\textbf{2, 4 or 5}$ & $\textbf{3, 6 or 7}$ & & $\textbf{0, 1, 9}$ & $\textbf{2, 4 or 5}$ & $\textbf{3, 6 or 7}$\\
			\cline{1-4}
			\cline{6-8}
			\cline{10-12}
		 \multirow{3}{*}{\centering{\textbf{TR-QNet}}} & 4 & & $\textbf{0.815}$ & & $\textbf{0.738}$ & $0.723$ & $\textbf{0.746}$ & & $0.707$ & $0.712$ & $0.723$\\	
                & 6 & & $0.802$ & & $\textbf{0.741}$ & $\textbf{0.736}$ & $\textbf{0.739}$ & & $\textbf{0.719}$ & $\textbf{0.725}$ & $\textbf{0.731}$\\
			& 8 & & $0.785$ & & $0.709$ & $0.717$ & $0.716$ & & $\textbf{0.718}$ & $0.707$ & $0.713$\\
          \hline
    VQTN & 6 & & $\textbf{0.811}$ & & $0.714$ & $0.701$ & $0.698$ & & $0.688$ & $0.682$ & $0.694$\\
          \hline
    VTNN & N/A & & $0.774$ & & $0.689$ & $0.684$ & $0.679$ & & $0.613$ & $0.639$ & $0.657$\\
    \hline
		    \label{tab3}
		\end{tabular}
	\end{center}
\end{table}

\section{Discussions}
\label{discuss}

The experimental results reported in the manuscript show that the proposed TR-QNet model outperforms its quantum and classical counterparts for binary classification and multi-class (ternary) classification in test datasets in the given experimental settings. This is due to the fact that the proposed TR-QNet is capable of modulating classification tasks by substituting the trainable weight matrices of the fully connected dense layers of standard TNN with Quan-TR, and hence, TNN acts as an efficient encoding tool, especially for large image features with minimal loss of information from the input images. The VQC-based training algorithm resembling DMRG~\cite{white1992} enables a straightforward entanglement of the entanglement spectrum of the MPO's~\cite{panaki2021} trainable weights, thereby facilitating a lucid comprehension of the correlations within the parameters of TN layers. For efficient training of the proposed Quan-TR model, we have presented a novel entanglement-aware training technique relying on hybrid classical-quantum algorithms and stochastic gradient-descent updates. This approach operates on a condensed parameter subspace obtained from the tensorization of trainable weights, leading to faster convergence and promising results. \\
Moreover, our implementation enables the creation of hybrid architectures that combine TN layers, dense layers, and Quan-TR to create true instances of deep learning models. 
Moreover, it is worth noting that the multi-layer design of Quan-TR within our proposed TR-QNet has the potential to produce the cascading effect of entanglement between the neuronal inputs and their outputs. Our results indicate that the classical TNNs with DMRG-like training and Quan-TR methods work accurately and efficiently for data and image classification tasks. The direct access to the singular values throughout the virtual dimensions of the trainable MPOs of TN layers provided by the DMRG-like training method and tensor ring optimized variational learning algorithm is crucial as it enables the computation of a measure of entanglement (correlation) between the features and model parameters. Since more qubits signify a bigger Hilbert space to parametrize the input data~\cite{biamonte2017}, we see a general pattern of increasing classification accuracy with qubit count from $4$ to $6$. However, Further increasing to $8, 10$ and $12$ qubits resulted in a substantial decrease in accuracy for the proposed TR-QNet, probably as a result of over-parametrization~\cite{larocca2023} and barren plateaus~\cite{mcclean2018}. Due to the additional non-linearity caused by the truncated singular value decomposition over the MPOs and two-qubit gate transformations, we also notice that in the case of the Iris dataset, TR-QNet significantly outperforms the VQTN, QCNN, and TR-VQC with full quantum state information and classically simulated VTNN. 
Eight-qubit circuit topologies are used in a series of studies utilizing different rankings to examine the impact of tensor ring rank on the performance of TR-QNet as it yields optimal results regarding input qubit counts.\\
However, the proposed TR-QNet model for multi-class image classification has achieved a comparable level of precision, primarily due to the inherent challenges faced by the slow convergence of Quan-TR. Hence, even though its promising performance is exhibited on relatively smaller datasets, the proposed TR-QNet is restricted due to the inherent difficulties in scaling and time-intensive training of Quan-TR. Nevertheless, TR-QNet has achieved higher accuracy in binary classification tasks when compared with its quantum and classical counterparts. Our method paves the way for developing novel deep neural network representations of a quantum state. It serves as a useful tool for investigating the expressive potential of quantum neural states. We aim to develop an efficient TR-QNet model comprising an optimized Quan-TR with fewer hyper-parameters. The total number of parameters in TNN is estimated as $O(\mathcal{N}_T d_T n + (n-2)d_T^3)$~\cite{ose2011}. In the case of Quan-TR, the computational complexity is $O(\mathcal{N}_q d_q)$ as each calculation of a single or two-qubit gate in the proposed Quan-TR is $O(1)$~\cite{peddi2023}.

\section{Conclusion}
\label{conlud}

In line with the impressive advances in quantum machine learning, the proposed TR-QNet framework offers an improvement over fully classical TNN and has been developed as a proof of concept using hybrid classical-quantum algorithms for better training strategies for TNN. In this paper, we have investigated the benefits of a Tensor Ring optimized variational Quantum learning classifier (Quan-TR) to find a better optimization strategy for TR-QNet, which exploits the entanglement inherent between qubits. The experimental results on the test datasets using the proposed TR-QNet model show its efficiency over the quantum and classical counterparts in binary and multi-class classification. Moreover, the experimental results demonstrate the efficacy of the proposed TR-QNet in various settings, which is crucial for data classification and image recognition in noisy intermediate-scale quantum (NISQ) devices. Consequently, our TR-QNet model is a strong contender for deep learning and can revolutionize the studies in quantum machine learning. \\
However, it remains to investigate the current TR-QNet architecture for deep convolutional neural networks and their training algorithms for regression and classification, which can be deployed immediately in near-term quantum devices. Authors are engaged in this direction. 

\section*{Appendix}
\label{appendix}
\section*{Convergence Analysis of TR-QNet} 
\label{PQIS-Net:Conv}

Due to NISQ's limitations, classical simulators are now being utilized to optimize and update parameters and feed them back to Tensor Neural Network (TNN) and Quan-TR separately until convergence conditions are reached. Hence, we have used cross-entropy loss to update the parameters.
 The loss function ($\mathcal{L}_{\theta}$) is derived with the hyper-parameters $\theta$ of the proposed TR-QNet model as
\begin{equation}
\begin{split}
   \operatorname*{argmin}_\theta \mathcal{L}_{\theta} = \sum_{j}^{\mathcal{N}_q} [t^j \log \overline{f}(\mathcal{O}_{TN}^j) + (1-t^j) \log \{1-\overline{f}(\mathcal{O}_{TN}^j)\}]
   \end{split}
   \label{eq:loss_func}
\end{equation}
$t^j$ corresponds to a target output, $\mathcal{N}_q$ is the number of qubits in Quan-TR and $\overline{f}(\mathcal{O}_{TN}^j)$ is the average outcome on quantum measurement of a qubit $j$ concerning the network hyper-parameter set $\theta$ as evaluated in the following subsection as follows. 
\begin{equation}
\begin{split}
    f(y_j (\theta), t^j) =\sum_{j}^{\mathcal{N}_q} f((\langle 0|\psi^\dagger (\mathcal{O}_{TN}^j)\mathcal{U}^\dagger (\theta) y_j \mathcal{U} (\theta)\psi (\mathcal{O}_{TN}^j)|0\rangle), t^j)
    \end{split}
\end{equation}
where, $y_j(\theta) \in \{\overline{\lambda_j}\}$ and $t^j$ corresponds to a target output and the preprocessed data from the TNN layer, denoted as $\mathcal{O}_{TN}^i$, is transformed into a quantum state represented by $|\psi(\mathcal{O}_{TN}^i)\rangle$.\\
In order to train the proposed Quan-TR model, the gradient of the loss function is evaluated as follows:
\begin{equation}
\begin{split}
\frac{\delta \overline{f}^\iota(\mathcal{O}_{TN}^j)}{\theta^j} = \langle 0|\psi^\dagger (\mathcal{O}_{TN}^j)\frac{\delta\mathcal{U}^\dagger (\theta)}{\delta \theta_j} y_j \mathcal{U} (\theta)\psi (\mathcal{O}_{TN}^j)|0\rangle +  \langle 0|\psi^\dagger (\mathcal{O}_{TN}^j)\mathcal{U}^\dagger (\theta) y_j \frac{\delta\mathcal{U} (\theta)}{\delta \theta_j}\psi (\mathcal{O}_{TN}^j)|0\rangle\\ = \langle 0|\psi^\dagger (\mathcal{O}_{TN}^j)\mathcal{U}_1^\dagger (\theta_1)\cdot\frac{\delta\mathcal{U}_j^\dagger (\theta_j)}{\delta \theta_j}\cdot \mathcal{U^\dagger}_n (\theta_n)y_j \mathcal{U} (\theta)\psi (\mathcal{O}_{TN}^j)|0\rangle + \\ \langle 0|\psi^\dagger (\mathcal{O}_{TN}^j)\mathcal{U}^\dagger (\theta) y_j \mathcal{U}_n (\theta_n)\cdot\frac{\delta\mathcal{U}_j (\theta_j)}{\delta \theta_j}\cdot\mathcal{U}_1 (\theta_1)\psi (\mathcal{O}_{TN}^j)|0\rangle
\end{split}
\label{eq:3}
\end{equation}
where, $\mathcal{U}_j (\theta_j) = e^{-i\theta_j\psi (\theta^j)}$. The global phase has no direct bearing on the results of the measurement, and hence, we disregard the global phase. Now, rotation gates can be written as follows.
\begin{equation}
\begin{split}
       \frac{\delta \psi (\theta, \mathcal{O}_{TN})}{\delta \theta^j} = \frac{1}{2} \psi (\theta + \frac{\pi}{2}, \mathcal{O}_{TN})
        \frac{\delta \psi^\dagger (\theta, \mathcal{O}_{TN})}{\delta \theta^j} = \frac{1}{2} \psi (\theta - \frac{\pi}{2}, \mathcal{O}_{TN})
\end{split}
\label{eq:4}
\end{equation}
Substituting Equation~\ref{eq:3} by Equation~\ref{eq:4}, we obtain as follows.
\begin{equation}
\begin{split}
\frac{\delta \overline{f}^\iota(\mathcal{O}_{TN}^j)}{\theta^j}= \langle 0|\psi^\dagger (\mathcal{O}_{TN}^j)\mathcal{U}_1^\dagger (\theta_1)\cdot\frac{\delta\mathcal{U}_j^\dagger (\theta_j)}{\delta \theta_j}\cdot \mathcal{U^\dagger}_n (\theta_n)y_j \mathcal{U} (\theta)\psi (\mathcal{O}_{TN}^j)|0\rangle + \\ \langle 0|\psi^\dagger (\mathcal{O}_{TN}^j)\mathcal{U}^\dagger (\theta) y_j \mathcal{U}_n (\theta_n)\cdot\frac{\delta\mathcal{U}_j (\theta_j)}{\delta \theta_j}\cdot\mathcal{U}_1 (\theta_1)\psi (\mathcal{O}_{TN}^j)|0\rangle\\
=\frac{1}{2}\{-\langle 0|\psi^\dagger (\mathcal{O}_{TN}^j)\mathcal{U}_1^\dagger (\theta_1)\cdot\frac{\delta\mathcal{U}_j^\dagger (\theta_j-\frac{\pi}{2})}{\delta \theta_j}\cdot \mathcal{U^\dagger}_n (\theta_n)y_j \mathcal{U} (\theta)\psi (\mathcal{O}_{TN}^j)|0\rangle + \\ \langle 0|\psi^\dagger (\mathcal{O}_{TN}^j)\mathcal{U}^\dagger (\theta) y_j \mathcal{U}_n (\theta_n)\cdot\frac{\delta\mathcal{U}_j (\theta_j+\frac{\pi}{2})}{\delta \theta_j}\cdot\mathcal{U}_1 (\theta_1)\psi (\mathcal{O}_{TN}^j)|0\rangle\}\\
= \frac{1}{2}\{\langle 0|\psi^\dagger (\mathcal{O}_{TN}^j)\mathcal{U}_{-}^\dagger[i\psi_j] \mathcal{U}_{+}^\dagger y_j \mathcal{U} (\theta)\psi (\mathcal{O}_{TN}^j)|0\rangle -  \langle 0|\psi^\dagger (\mathcal{O}_{TN}^j)\mathcal{U}^\dagger (\theta) y_j \mathcal{U}_{+}[-i\psi_j]\mathcal{U}_{-}\psi (\mathcal{O}_{TN}^j)|0\rangle \}\\
= \frac{1}{2} \Psi_{\theta_{+}} (\psi (\mathcal{O}^j_{TN})) -\frac{1}{2} \Psi_{\theta_{-}} (\psi (\mathcal{O}^j_{TN}))
\end{split}
\end{equation}
For the rotation gates $\mathcal{R}_y (\omega_y)$ and $\mathcal{R}_z(\omega_z)$ of Quan-TR in TR-QNet, the angle of rotation [variational parameter ($\theta$)] is $\omega_y$ and $\omega_z$, respectively. The rotation gates $\mathcal{R}_y (\omega_y)$ and $\mathcal{R}_z(\omega_z)$ of Quan-TR operate the qubits $|\psi_y\rangle $ and $|\psi_z\rangle$ as follows.
\begin{equation}
|\psi_y(\iota+1)\rangle=\left(
\begin{array}{cc}
\cos \triangle \omega_y(\iota) & -\sin \triangle \omega_y(\iota) \\
\sin \triangle \omega_y(\iota) & \cos \triangle \omega_y(\iota) \\
\end{array}
\right)|\psi_y(\iota) \rangle
\end{equation}
\begin{equation}
|\psi_z(\iota+1)\rangle=\left(
\begin{array}{cc}
\exp (-j\triangle \omega_z(\iota)) & 0 \\
     0 &  (-j\triangle \omega_z(\iota))\\
\end{array}
\right)|\psi_z(\iota) \rangle
\end{equation}
 where,
 \begin{equation}
 \omega_y(\iota+1)=\omega_y(\iota) + \triangle \omega_y(\iota) 
 \label{eq:alpha}
 \end{equation}
 and
 \begin{equation}
 \omega_z(\iota+1)=\omega_z(\iota) + \triangle \omega_z(\iota) 
 \label{eq:beta}
 \end{equation}
 For the quantum layers in Quan-TR at epoch, $\iota$, Equations~\ref{eq:alpha} and~\ref{eq:beta} measure the change in the phase or angles $\triangle \omega_y(\iota) $ and $\triangle \omega_z(\iota) $, respectively. Let us Consider 
\begin{equation}
\mathcal{C}(\iota) =\omega_y(\iota) - \overline {\omega_y(\iota)}
\end{equation}
\begin{equation}
\mathcal{D}(\iota) =\omega_z(\iota) - \overline {\omega_z(\iota)}
\end{equation}
and
\begin{equation}
\mathcal{R}(\iota) =\omega_y(\iota+1)-\omega_y(\iota) =\mathcal{C}(\iota+1)-\mathcal{C}(\iota) 
\end{equation}
\begin{equation}
\mathcal{S}(\iota) =\omega_z(\iota+1)-\omega_z(\iota) =\mathcal{D}(\iota+1)-\mathcal{D}(\iota) 
\end{equation}
The optimal phases or angles are therefore $\overline \omega_y(\iota) $ and $\overline \omega_z(\iota) $ for the rotation gates $\mathcal{R}_y (\omega_y)$ and $\mathcal{R}_z(\omega_z)$, respectively. \\
In order to update the weights in TNN, a gradient of the bond tensors with respect to the loss ($\mathcal{L}_\theta$), $\mathcal{B}^{j,j+1}$, is obtained by defining $\overline{f}(\mathcal{O}_{TN}^j)=\mathcal{T}\mathcal{B}$, where $T$ represents the contraction of every tensor in the TNN other than the bond tensor $\mathcal{B}$.\\
When considering $\mathcal{B}^j(\iota)$, $\omega_y ^j(\iota)$ and $\omega_z^j(\iota)$, the loss function $\mathcal{L}_\theta (\mathcal{B},\omega_y,\omega_z)$ is differentiated as follows:
\begin{equation}
\begin{split}
 \frac{\partial \mathcal{L}_\theta (\mathcal{B},\omega_y,\omega_z)}{\partial \mathcal{B}^j(\iota)} = \sum_{j=1}^{\mathcal{N}_q} \frac{\partial \overline{f}^\iota(\mathcal{O}_{TN}^j)}{\mathcal{B}^j(\iota)} \left[\frac{t^j}{\overline{f}^\iota(\mathcal{O}_{TN}^j)} - \frac{t^j-1}{1-\overline{f}^\iota(\mathcal{O}_{TN}^j)} \right]\\
 = \sum_{j=1}^{\mathcal{N}_q} \mathcal{T}^j(\iota) \left[\frac{t^j}{\overline{f}^\iota(\mathcal{O}_{TN}^j)} - \frac{t^j-1}{1-\overline{f}^\iota(\mathcal{O}_{TN}^j)} \right]
 \end{split}
 \end{equation}
 Hence, the change in the bond tensor designated as $\triangle \mathcal{B}^j(\iota)$ is evaluated as follows. 
\begin{equation}
    \triangle \mathcal{B}^j(\iota) = -\gamma(\iota)\frac{\partial \mathcal{L}_\theta (\mathcal{B},\omega_y,\omega_z)}{\partial \mathcal{B}^j(\iota)}
\end{equation}
Here, $\gamma(\iota)$ is a learning rate in the gradient descent procedure for updating the bond tensors in TN layers. 
\begin{equation}
\begin{split}
 \frac{\partial \mathcal{L}_\theta (\mathcal{B},\omega_y,\omega_z)}{\partial \omega_y^j(\iota)} = \sum_{j=1}^{\mathcal{N}_q} \frac{\partial \overline{f}^\iota(\mathcal{O}_{TN}^j)}{\omega_y^j(\iota)} \left[\frac{t^j}{\overline{f}^\iota(\mathcal{O}_{TN}^j)} - \frac{t^j-1}{1-\overline{f}^\iota(\mathcal{O}_{TN}^j)} \right]
\end{split}
 \end{equation}
 \begin{equation}
\begin{split}
 \frac{\partial \mathcal{L}_\theta (\mathcal{B},\omega_y,\omega_z)}{\partial \omega_y^j(\iota)} =\sum_{j=1}^{\mathcal{N}_q} \frac{\partial \overline{f}^\iota(\mathcal{O}_{TN}^j)}{\omega_y^j(\iota)} \left[\frac{t^j}{\overline{f}^\iota(\mathcal{O}_{TN}^j)} - \frac{t^j-1}{1-\overline{f}^\iota(\mathcal{O}_{TN}^j)} \right]
\end{split}
 \end{equation}
 Here, parameter shift techniques are used to evaluate the gradient of the Quan-TR parameters $\omega_y$ and $\omega_z$~\cite{mitarai2018, li2017, huang2021} as follows. 
\begin{equation}
\frac{\partial \overline{f}^\iota(\mathcal{O}_{TN}^j)}{\omega_y^j(\iota)} = \frac{1}{2}\left[\Psi^{\iota+1}_{\omega_y+ \frac{\pi}{2}}(\psi (\mathcal{O}^j_{TN}))-\Psi^\iota _{\omega_y- \frac{\pi}{2}}(\psi (\mathcal{O}^j_{TN}))\right]
 \end{equation}
 and 
\begin{equation}
\frac{\partial \overline{f}^\iota(\mathcal{O}_{TN}^j)}{\omega_z^j(\iota)} = \frac{1}{2}\left[\Psi^{\iota+1}_{\omega_z+ \frac{\pi}{2}}(\psi (\mathcal{O}^j_{TN}))-\Psi^\iota _{\omega_z- \frac{\pi}{2}}(\psi (\mathcal{O}^j_{TN}))\right]
 \end{equation}
where, with rotation angles $\omega_y^j(\iota) $ and $\omega_z^j(\iota) $, respectively, $\Psi(\iota) _{\omega_y\pm \frac{\pi}{2}}(\psi (\mathcal{O}^j_{TN}))$ and $\Psi(\iota)_{\omega_z\pm \frac{\pi}{2}}(\psi (\mathcal{O}^j_{TN}))$ are the measured qubit $\psi (\mathcal{O}^j_{TN})$.
The changes in phase or angles are designated as $\triangle \omega_y^j(\iota)$ and $\triangle \omega_z^j(\iota)$, respectively, for the rotation gate used to update the qubits. The rotation angles are then modified using the formula below.
\begin{equation}
\triangle \omega_y^j(\iota) =-\nu(\iota) \{\frac{\partial \overline{f}^\iota(\mathcal{O}_{TN}^j)}{\omega_y^j(\iota)}\}
\end{equation}
\begin{equation}
\triangle \omega_z^j(\iota) =-\mu (\iota) \{\frac{\partial \overline{f}^\iota(\mathcal{O}_{TN}^j)}{\omega_z^j(\iota)}\}
\end{equation}
Here, $\nu(\iota) $ and $\mu (\iota) $ are the learning rates in the gradient descent procedure for updating the rotation angles.
 
 \section{Data availability}
\label{data}
 The Iris dataset~\cite{iris_2021}, MNIST\cite{lecun1998}, and CIFAR-10~\cite{alexnet2012} datasets can be found in the following links:\url{https://archive.ics.uci.edu/dataset/53/iris}, \url{http://yann.lecun.com/exdb/mnist/}, and \url{https://www.cs.toronto.edu/~kriz/cifar.html}, respectively.
 
\section{Code Availability \& Description}
\label{code}
The PyTorch implementation of TR-QNet is available on Github: \url{https://github.com/konar1987/TR-QNet/}.

\section{Acknowledgements}
This work was partially supported by the Fulbright-Nehru Visiting Researcher Grant $\#2858 FNPDR/2022$.

\bibliographystyle{unsrt}
\bibliography{Deb.bib}

\end{document}